\documentclass{elsart}
\usepackage{graphicx,amssymb,lineno}

\def\pa{\partial}

\renewcommand{\theequation}{\thesection.\arabic{equation}}
\newcommand{\initiate}{\setcounter{equation}{0}}

\newcommand{\beq}{\begin{equation}}
\newcommand{\eeq}{\end{equation}}
\newcommand\be{\begin{equation} }
\newcommand\bea{\begin{eqnarray}}
\newcommand\ee{\end{equation}}
\newcommand\eea{\end{eqnarray}}

\def\Tr{{\rm Tr}\,}

\usepackage{epsfig}

\newlength{\myVSpace}
\def\endtitle{\par\end{quotation}\vskip3.5in minus2.3in\newpage}

\def\m{\mu}          \def\n{\nu}
       
       \def\r{\rho}
\def\s{\sigma}

         \def\G{\Gamma}

\def\ci{{\cal I}}

      \def\cv{{\cal V}}

\begin{document}

\begin{frontmatter}
\title{Non-commutative SU(N) gauge theories and asymptotic freedom}

\author[bg]{D. Latas}\ead{latas@phy.bg.ac.yu},
\author[bg]{V. Radovanovi\'c}\ead{rvoja@phy.bg.ac.yu},
\author[zg]{J. Trampeti\'c}\ead{josipt@rex.irb.hr}
\address[bg]{Faculty of Physics, University of Belgrade, \\
P.O.Box 368, 11001 Belgrade, Serbia}
\address[zg]{Rudjer Bo\v{s}kovi\'{c} Institute, Theoretical Physics Division,
   \\     P.O.Box 180, 10002 Zagreb, Croatia}

\begin{abstract}

In this paper we analyze the one-loop renormalization of the
$\theta$-expanded $\rm SU(N)$ Yang-Mills theory. We show that the
{\it freedom parameter} $a$, key to renormalization, originates
from
 higher order non-commutative gauge interaction,
represented by a higher derivative term
$\,b\,h\,\theta^{\mu\nu}\widehat F_{\mu\nu}\star\widehat
F_{\rho\sigma}\star\widehat F^{\rho\sigma}$.
 The renormalization
condition fixes the allowed values of the  parameter $a$ to
one of the two solutions: $a=1$ or $a=3$, i.e. to $b=0$ or
to $b=1/2$, respectively.
When the higher order interaction is switched on,
($a=3$), pure non-commutative SU(N) gauge theory
at first order in $\theta$-expansion becomes one-loop renormalizable
for various representations of the gauge group.
We also show that, in the case $a=3$ and   the adjoint representation
of the gauge fields,
the non-commutative deformation parameter
$h$ has to be renormalized and it is asymptotically free.

\end{abstract}


\end{frontmatter}

\section{Introduction}
\label{intro}
For some years it was believed that field theories defined on
non-commutative (NC) Minkowski space were not renormalizable. Namely,
if non-commutativity is canonical,
\begin{equation}
[\widehat x^\mu,\widehat x^\nu ] =\mathrm{i}h\theta^{\mu\nu}={\rm
const}, \label{Mink}
\end{equation}
then the algebra generated by the coordinates $\widehat x^\mu$,
i.e. non-commutative Min\-kow\-ski space and the fields  on it can
be represented by the algebra of functions on the ordinary ${\bf
R}^4$ with the Moyal-Weyl product instead of the usual
multiplication
\begin{equation}
\widehat\phi (x)\star \widehat\psi (x) =
e^{\frac{\mathrm{i}}{2}h\theta^{\m\n}\frac{\pa}{\pa
x^\m}\frac{\pa}{ \pa y^\n}}\widehat \phi (x)\widehat \psi
(y)|_{y\to x} . \label{moyal}
\end{equation}
Furthermore, the integral can be defined straightforwardly and has
the trace property. Thus  one can formulate field theories with the
action and the variational principle. However, in the quantization
of these $\theta$-unexpanded theories (e.g., $\phi^4$) one, as a
rule, meets the obstruction to renormalizability: the UV/IR mixing
\cite{Martin:1999aq,Minwalla:1999px,Matusis:2000jf}. 

In (\ref{Mink}) the non-commutative deformation parameter $h$ has dimension
$length^2$ or $energy^{-2}$, and can also be written as
$h=1/\Lambda^2_{\rm NC}$, where the $\Lambda_{\rm NC}$ represents
the scale of non-commutativity.

Gauge theories can  be extended to a non-commutative setting in
different ways. In our model, the classical action is obtained via a
two-step procedure. First, the action of the non-commutative
Yang-Mills (NCYM) theory is equipped with a star-product carrying
information about the underlying non-commutative manifold, and,
second, the star-product and non-commutative fields are expanded in
the non-commutativity parameter $h\theta$ using the Seiberg-Witten
(SW) map \cite{Seiberg:1999vs}. In this approach
\cite{Wess,Buric:2005xe,Buric:2006wm,Martin:2006gw},
non-commutativity is treated perturbatively.
The major advantage is that models with any gauge
group and any particle content can be constructed
\cite{Wess,Calmet:2001na,Blazenka,Aschieri:2002mc,Goran}, so we can
construct the generalization of the standard model (SM), too. The
action is gauge invariant; furthermore, it has been proved that the
action is anomaly free whenever its commutative counterpart is also
anomaly free \cite{Brandt:2003fx}.

In this paper, which is a continuation of two recent papers,
\cite{Buric:2005xe} and \cite{Buric:2006wm}, we analyze the
renormalizability property of Yang-Mills theory on non-commutative
space, where we confine ourselves to the $\theta$-expanded NC SU(N)
gauge theory. Commutative  gauge symmetry is the underlying symmetry
of the theory and is present in each order of the
$\theta$-expansion. Non-commutative symmetry, on the other hand,
exists only in the full theory, i.e. after the summation.

There are a number of versions of the non-commutative standard
model (NC\-SM) in the $\theta$-expanded approach,
\cite{Calmet:2001na,Blazenka,Aschieri:2002mc,Goran}. The argument
of renormalizability was previously included in the construction
of field theories on non-co\-mmu\-ta\-ti\-ve Minkowski space
producing not only the one-loop renormalizable model
\cite{Buric:2005xe}, but the model containing one-loop quantum
corrections free of divergences \cite{Buric:2006wm}, contrary to
previous results
\cite{Bichl:2001cq,Wulkenhaar:2001sq,Grimstrup:2002af,Maja}. This
`good' behavior of the $\theta$-expanded non-commutative SM gauge
theory is our primary motivation to re-examine one-loop
renormalizability aspect of the pure NC SU(N) gauge sector. We
shall perform the analysis at first order in $\theta$, and for the
fundamental representations of the matter field. Phenomenological
consequences of this investigation are certainly important
\cite{Goran,Josip,Ohl:2004tn}.

The plan of the paper is the following. In Section 2 we briefly
review the ingredients of the SW freedom  and construct the
higher-order Lagrangian term which renders one-loop
renormalizability of the non-commutative theories at  first order
in $\theta$. In Section 3 the one-loop renormalizability of the
pure NC SU(N) gauge theory is worked out. Section 4 is devoted to
the ultraviolet asymptotic behavior of NC SU(N) gauge theory. The
discussion of the results and the concluding points are given in
Section 5.

\initiate \section{NC SU(N) gauge sector effective action}
\label{sect2}

According to \cite{Wess}, the NC parameter $\widehat \Lambda$, the
NC vector potential $\widehat V_\mu$ and the corresponding NC
field strength $\widehat F_{\mu\nu}$ take their values in the
enveloping algebra of the Lie algebra of the gauge group. As in
ordinary theory, in the non-commutative case, symmetry is
localized by the NC vector potential $\widehat V_\mu$ and the NC
field strength $\widehat F_{\mu\nu}$
\begin{equation}
\widehat F_{\mu\nu} = \pa_\mu\widehat V_\nu - \pa_\nu \widehat
V_\mu - \mathrm{i}(\widehat V_\mu\star\widehat V_\nu - \widehat
V_\nu\star\widehat V_\mu ). \label{f}
\end{equation}
We start by solving the gauge field transformation closure
condition \cite{Wess} order by order in the  parameter $h$. The
solutions up to the first order for the vector field and the field
strength read
\begin{eqnarray}
\widehat V_\mu(x) &=&V_\mu(x) -\frac 14 h \theta ^{\s\r}\left\{
V_\s(x), \pa _\r V_\m(x) +F_{\r \m}(x)\right\}+ \dots
\label{SW}\\
{\widehat F}_{\mu\nu}(x)&=&F_{\mu\nu} +
\frac{1}{4}h\theta^{\sigma\r} \left(2\{F_{\mu\sigma}, F_{\nu\r}\}
-\{V_{\sigma} ,(\partial_{\r}+{\cal D}_{\r})F_{\mu\nu} \} \right)
+ \dots
\label{fields1}
\end{eqnarray}
The non-Abelian field strength and the covariant derivative are
defined in the usual way
$F_{\mu\nu}=\partial_{\mu}V_{\nu}-\partial_{\nu}V_{\mu}-
\mathrm{i}[V_{\mu},V_{\nu}]$ and ${\cal
D}_{\mu}=\partial_{\mu}-\mathrm{i}[V_{\mu},\quad]$. The relations
(\ref{SW}-\ref{fields1}) between non-commutative and commutative
gauge symmetries are known as the Seiberg-Witten maps,
\cite{Seiberg:1999vs}. For zero non-commutativity, $\widehat V_\m$
and $\widehat F_{\m\n}$ reduce to the usual vector potential $V_\mu$
and the field strength $F_{\m\n}$. The SW map, which fulfills a
number of requirements (hermiticity, non-uniqueness etc), leading to
a physically acceptable theory, was discussed extensively in
\cite{Martin:2005vr,Martin:2005jy}.

Clearly, the solution (\ref{SW}) is not unique. Non-uniqueness is
given by  the transformation
\begin{equation}
{\widehat V}_{\mu} \to {\widehat V}_{\mu} +  X_\mu ,\quad
{\widehat F}_{\mu\nu} \to{\widehat F}_{\mu\nu} + D_\mu X_\nu -
D_\nu X_\mu,
\label{nonuniq}
\end{equation}
which one understands as freedom to define the physical fields
$V_\mu$ and the field strengths $F_{\mu\nu}$\footnote{ The
transformation  (\ref{nonuniq}) introduced in \cite{Buric:2006wm}
in fact does not produce new terms in the gauge field lagrangian
as we claimed,  \cite{oho!}. However, this does not spoil  the
main result (3.21) of Ref. \cite{Buric:2006wm}, that is the
renormalizability and finiteness of the 1-loop divergent terms in
the nmNCSM gauge sector for the choice of the parameter $a=3$.  In
order to obtain terms necessary for renormalization, we have to
introduce higher-order NC gauge interactions as we will see later
on. }.

The usual, or minimal NC SU(N) gauge theory action
\begin{equation}
S_g=-\frac{1}{2}\Tr \int \mathrm{d}^4x\widehat F_{\m\n}\star\widehat
F^{\m\n} , \label{action}
\end{equation}
expanded in the deformation parameter $h$
using SW map (\ref{fields1}) reads
\begin{equation}
S_g^1 =\Tr\int \mathrm{d}^4 x\, \left[-\frac{1}{2}F_{\m\n}F^{\m\n}+h
\theta^{\m\n}  \left( \frac{1}{4}
F_{\m\n}F_{\r\s}-F_{\m\r}F_{\n\s} \right)F^{\r\s}\right].
\label{act}
\end{equation}
One can however generalize this action introducing the terms of
higher order in NC field strengths. Due to the Lorentz structure
there are only two third-order interaction terms, producing the
generalized NC gauge
 action
\begin{eqnarray}
S_{\star} &=&\Tr\int \mathrm{d}^4x\left[-\frac{1}{2}{\widehat
F_{\mu\nu}}\star{\widehat F^{\mu\nu}} +h\theta^{\mu\nu}
\left(b\widehat F_{\mu\nu}\star\widehat F_{\rho\sigma}+
c\widehat F_{\mu\rho}\star\widehat
F_{\nu\sigma}\right) \star\widehat F^{\rho\sigma}\right],
\nonumber\\
\label{GenAction}
\end{eqnarray}
with still unspecified constants $b$ and
$c$. We will not consider here the terms of fourth and higher order in
NC field strengths. Let us stress
once again that the action $S_{\star}$ is invariant under the NC
gauge transformation. The constants $b$ and
$c$ are going to be  restricted   further.

Substituting the SW map for NC field strength (\ref{fields1}) into
the action (\ref{GenAction}), we obtain
\begin{eqnarray}
S^1_{\star}&=&\Tr\int \mathrm{d}^4x\left[\, -\frac{1}{2}F_{\mu\nu}
F^{\mu\nu} +h\theta^{\mu\nu}\, \left(\frac{1}{4}
F_{\mu\nu}F_{\rho\sigma}-
F_{\mu\rho}F_{\nu\sigma}\right)F^{\rho\sigma}\right.
\nonumber\\
&&\,\left. \phantom{\star{\widehat
F^{\mu\nu}}\theta^{\mu\nu}\frac{a}{2}} +h\theta^{\mu\nu}\,
\left(b F_{\mu\nu} F_{\rho\sigma}+ c
F_{\mu\rho}F_{\nu\sigma}\right)F^{\rho\sigma}\right],
\nonumber\\
&=&\Tr\int \mathrm{d}^4x\left[\,-\frac{1}{2} F_{\mu\nu} F^{\mu\nu}
\right.\nonumber \\ &&\,\left. \phantom{\star{\widehat
F^{\mu\nu}}\theta^{\mu\nu}\frac{a}{2}} +h\theta^{\mu\nu}\,
\left((\frac14+b)F_{\mu\nu}F_{\rho\sigma}+
(c-1)F_{\mu\rho}F_{\nu\sigma}\right)F^{\rho\sigma}\right]\,.
\nonumber\\
\label{GA}
\end{eqnarray}
The above-given action  is invariant under the classical gauge
transformation. In order to obtain new contributions from
$S^1_\star$ the constant $c$ has to respect requirement  $c\not=
1$. In the expression (\ref{GA}) we are free to choose $c=0$, and
then, via simple redefinition,
\begin{equation}
1+4b =a\,,
\label{a}
\end{equation}
we arrive at the desired  action $S$  to first order in
 non-commutative deformation parameter $h$:
\begin{equation}
S =\Tr\int \mathrm{d}^4x\,\left[-\frac{1}{2}
F_{\m\n}F^{\m\n}+h\theta^{\m\n}\, \left(
\frac{a}{4} F_{\m\n}F_{\r\s}-F_{\m\r}F_{\n\s} \right)F^{\r\s}\right]\,.
\label{actioa}
\end{equation}
Here $a$ is an arbitrary real parameter, i.e. {\it the freedom
parameter} to be determined, as before \cite{Buric:2006wm}, from
the renormalizability requirement\footnote{It was found in
\cite{Buric:2005xe} that for $a=1$ the theory is renormalizable,
while in \cite{Buric:2006wm} renormalizability/finiteness of the
model required $a=3$.}.

From the expression (\ref{a}) it is clear that the $a$ dependence of the
gauge action $S$, (\ref{actioa}), crucial to obtain
renormalizability/finiteness of the nmNCSM, \cite{Buric:2006wm},
is arising from the inclusion of the higher order gauge
interaction term $(\,\theta^{\mu\nu}\widehat F_{\mu\nu}\star
\widehat F_{\rho\sigma}\star\widehat F^{\rho\sigma})$
into the action (\ref{GenAction}),
and from the implementation of SW map (\ref{fields1}).

Finally, it is important to
notice that the $h$-linear terms in (\ref{GA}) and (\ref{actioa})
depend on the representation of gauge fields: they are
proportional to the trace of the product of three group
generators.  Thus the non-commutative correction to the
gauge field action depends on the representations of fields in a
given theory.

To find the classical action, we follow \cite{Buric:2005xe}. For
non-Abelian vector fields,  from (\ref{actioa}) we obtain
\begin{eqnarray}
S_{\mathrm{NCYM}}&=&\int \mathrm{d}^4x \left[-\frac{1}{4}
F^a_{\mu\nu} F^{a\mu\nu}\right. \nonumber \\& &\phantom{\int
\mathrm{d}^4x\Big)}\left. + \frac{1}{4} h\theta^{\mu\nu} d^{abc}
\left( \frac{a}{4} F^a_{\mu\nu} F^b_{\rho\sigma} - F^a_{\mu\rho}
F^b_{\nu\sigma} \right) F^{c\rho\sigma} \right],
\label{m}
\end{eqnarray}
where $d^{abc}$ are totally symmetric coefficients of the SU(N)
group. This action, for $a=1$, corresponds to the classical action
$S_{\rm mNCSM}$, i.e. to Eq. (24) constructed in \cite{Blazenka}.
Here, for the general SU(N) gauge group, $a,b,c=1,\dots,N^2-1$ are
the group indices. Finally, note that in \cite{Buric:2006wm} the
action of the type (\ref{m}) was absent owing to the special
choice of the representation SU(3)$_\mathrm{C}$ group where the
coefficient $\kappa^{abc}_5$ was zero.

\initiate \section{One-loop renormalization}

To perform  the one-loop renormalization of the NC SU(N) gauge part
action (\ref{actioa}), we apply, as before
\cite{Buric:2005xe,Buric:2006wm}, the background-field method
\cite{'tHooft:1973us,PS}.  As we have already explained the
details of the method in \cite{Maja}, here we only discuss the
points needed for this computation. The main contribution to the
functional integral is given by the Gaussian integral. However,
technically, this is achieved by splitting the vector potential
into the classical-background and the quantum-fluctuation parts,
that is, $\phi_V\to \phi_V + {\bf\Phi}_V$, and by computing the
terms quadratic in the quantum fields. In this way we determine
the second functional derivative of the classical action, which is
possible since our interactions (\ref{actioa}) and/or (\ref{m})
are of the polynomial type. The quantization is performed by the
functional integration over the quantum vector field ${\bf
\Phi}_V$ in the saddle-point approximation around the classical
(background) configuration $\phi _V$.

First, an advantage of the background-field method is that it
guarantees  covariance, as in  doing the path integral the local
symmetry of the quantum field ${\bf \Phi}_V$ is fixed, while the
gauge symmetry of the background field ${\phi}_V$ is manifestly
preserved.

Since we are dealing with gauge symmetry, our Lagrangian (\ref{m})
is singular owing to its invariance under the gauge group.
Therefore, a proper quantization of (\ref{m}) requires the
presence of the gauge fixing term $S_{\rm gf}[\phi]$ in the
one-loop effective action
\begin{equation}
\Gamma [\phi] = S_{\rm cl}[\phi] + S_{\rm gf}[\phi] + \Gamma^{(1)}
[\phi],\qquad S_{\rm gf}[\phi]=-\frac 12 \int
\mathrm{d}^4x(D_{\mu}{\bf \Phi}_V^{\mu})^2, \label{GS2}
\end{equation}
producing the standard result of the commutative part of our
action (\ref{actioa}). In $S_{\mathrm{gf}}$ from (\ref{GS2}) we
have chosen the Feynman gauge `$\alpha=1$'.

The one-loop effective part $\Gamma^{(1)} [\phi]$ is given by
\begin{equation}
\Gamma ^{(1)}[\phi] =\frac{\mathrm{i}}{2}\log\det
S^{(2)}[\phi]=\frac{\mathrm{i}}{2}\Tr\log S^{(2)}[\phi].
\label{gama1}
\end{equation}
In (\ref{gama1}), the $ S^{(2)}[\phi]$ is the second functional
derivative of the classical action \be S^{(2)}[\phi]= \frac
{\delta^2 S_{\mathrm{cl}}}{\delta\phi_{V_1} \delta\phi_{{V_2}}} .
\label{S2} \ee The structure of $S^{(2)}[\phi]$ is
\begin{equation}
S^{(2)}=\Box+N_1+N_2+T_2+T_3+T_4 , \label{Box}
\end{equation}
where $N_1$, $N_2$ are commutative vertices, while $T_2$, $T_3$,
$T_4$ are non-commutative ones. The indices denote the number of
classical fields.  The one-loop effective action computed by using
the background-field method is \bea \G^{(1)}_{\theta,2}&=&
\frac{\mathrm{i}}{2} \Tr \log \left(\ci + \Box ^{-1}
(N_1+N_2+T_2+T_3+T_4)\right)
\label{trlog}\\
&=& \frac{\mathrm{i}}{2}\sum_{n=1}^\infty \frac{(-1)^{n+1}}{n} \Tr
\left(\Box^{-1}N_1+\Box ^{-1}N_2+\Box
^{-1}T_2+\Box^{-1}T_3+\Box^{-1}T_4 \right)^n. \nonumber \eea For
dimensional reasons, the divergences in  $h\theta$-linear order
are all of the forms $h\theta F V^4$, $h\theta F^2 V^2$ and
$h\theta F^3$. Since the $h\theta$-3-, $h\theta$-4-, $h\theta$-5-
and $h\theta$-6-vertices obtain divergent contributions, from the
sum (\ref{trlog}) we need to  extract and compute only terms that
contain up to three external field strengths.

As  the conventions and the notation are the same as in
\cite{Buric:2005xe}, we only en\-coun\-ter and discuss the
intermediate and the final results.

Using the previously introduced notation, the vertices  read
\begin{eqnarray}
(N_1)^{ab\alpha\beta} & =& -2 \mathrm{i} (V^\mu)^{ab}
g^{\alpha\beta}
\partial_\mu ,
\label{N1} \\
(N_2)^{ab\alpha\beta} & =& -2 f^{abc} F^{c\alpha\beta} - (V^\mu
V_\mu)^{ab} g^{\alpha\beta} , \label{N2}
\end{eqnarray}
which are the same as in the commutative case. Non-commutative
vertices are
\begin{eqnarray}
(T_2)^{ab\alpha\beta} & =& \frac{h}{8} d^{abc} \left\{
\left[\left( a \theta^{\rho\sigma} F^{c}_{\rho\sigma}
g^{\alpha\nu} g^{\beta\mu} - 2(a-1) \theta^{\alpha\mu} F^{c
\beta\nu} + 4 {\theta^\alpha}_\rho
F^{c\beta\rho} g^{\mu\nu} \right.\right.\right. \nonumber \\
& &\left.\left.\phantom{1 d^{abc}d^b}+ 4 {\theta^\mu}_\rho
F^{c\nu\rho} g^{\alpha\beta}\right) -(\beta \leftrightarrow \nu)
\right]  \left.+ [\alpha\leftrightarrow\beta] \right\}\partial_\mu
\partial_\nu ,
\label{T2} \\
(T_3)^{ab\alpha\beta} & =& \frac{\mathrm{i}h}{4} \left\{ d^{acd}
\left[ -2a\theta^{\alpha\mu} (V_\nu)^{bc} F^{d\beta\nu} - 2a
\theta^{\beta\nu} (V_\nu)^{bc} F^{d\alpha\mu} \right.\right.
\nonumber \\
& & \phantom{\frac{i}{4} \{ d^{acd} \Big[ } - a
\theta^{\rho\sigma} (V^\mu)^{bc} F^d_{\rho\sigma} g^{\alpha\beta}
+ a \theta^{\rho\sigma} (V^\alpha)^{bc} F^d_{\rho\sigma}
g^{\beta\mu}
\nonumber \\
& & \phantom{\frac{i}{4} \{ d^{acd} \Big[ } - 2
{\theta^\alpha}_\rho (V_\nu)^{bc} F^{d\nu\rho} g^{\beta\mu} + 2
{\theta^\alpha}_\nu (V^\mu)^{bc} F^{d\beta\nu}
\nonumber \\
& & \phantom{\frac{i}{4} \{ d^{acd} \Big[ } + 2 \theta^{\mu\rho}
(V^\nu)^{bc} F^{d}_{\nu\rho} - 2 {\theta^{\mu}}_\rho
(V^\alpha)^{bc} F^{d\beta\rho} -2 {\theta^{\beta}}_\rho
(V^\alpha)^{bc} F^{d\mu\rho}
\nonumber \\
& & \phantom{\frac{i}{4} \{ d^{acd} \Big[ }  + 2
{\theta^{\beta}}_\rho (V^\mu)^{bc} F^{d\alpha\rho} + 2
{\theta^\nu}_\rho (V_\nu)^{bc} F^{d\mu\rho} g^{\alpha\beta}
\nonumber\\
& & \phantom{\frac{i}{4} \{ d^{acd} \Big[ }   +  2
\theta^{\alpha\beta} (V_\nu)^{bc} F^{d\mu\nu} + 2
\theta^{\alpha\nu} (V_\nu)^{bc} F^{d\beta\mu} + 2\theta^{\beta\mu}
(V_\nu)^{bc} F^{d\alpha\nu}
\nonumber \\
& & \phantom{\frac{i}{4} \{ d^{acd} \Big[ } \left.- 2
{\theta^\nu}_\rho (V_\nu)^{bc} F^{d\alpha\rho} g^{\beta\mu} +  2
\theta^{\mu\nu} (V_\nu)^{bc} F^{d\alpha\beta}\right]
\nonumber \\
& & \left.\phantom{{4} \{ d^{acd} \Big[ } - [a\leftrightarrow
b,\alpha\leftrightarrow\beta] \right\} \partial_\mu,
\label{T3} \\
(T_4)^{ab\alpha\beta} & =& \frac{h}{8} d^{cde} \left[ \left(-4a
\theta^{\alpha\rho} (V_\rho)^{ac} (V_\mu)^{bd} F^{e\beta\mu} -a
\theta^{\rho\sigma} (V^\mu)^{ac} (V_\mu)^{bd} F^{e}_{\rho\sigma}
g^{\alpha\beta} \right.\right.
\nonumber \\
& & \phantom{\frac{1}{8} d^{cde} \Big[ (}  +a \theta^{\rho\sigma}
(V^\beta)^{ac} (V^\alpha)^{bd} F^{e}_{\rho\sigma} -4
\theta^{\alpha\rho} (V^\beta)^{ad} (V^\mu)^{bc} F^{e}_{\mu\rho}
\nonumber \\ & & \phantom{\frac{1}{8} d^{cde} \Big[ (}   + 4
{\theta^\alpha}_\rho (V^\mu)^{ad} (V_\mu)^{bc} F^{e\beta\rho} + 4
{\theta^\nu}_\rho (V_\nu)^{ad} (V_\mu)^{bc} F^{e\mu\rho}
g^{\alpha\beta}
\nonumber \\
& & \phantom{\frac{1}{8} d^{cde} \Big[ (} + 2 \theta^{\alpha\beta}
(V_\mu)^{ad} (V_\nu)^{bc} F^{e\mu\nu} + 4 \theta^{\alpha\rho}
(V_\mu)^{ad} (V_\rho)^{bc} F^{e\beta\mu}
\nonumber \\
& & \phantom{\frac{1}{8} d^{cde} \Big[ (} \left. +
2\theta^{\rho\sigma} (V_\rho)^{ad} (V_\sigma)^{bc}
F^{e\alpha\beta} \right)+(a\leftrightarrow b\,,
\alpha\leftrightarrow\beta)
\nonumber \\
& & \phantom{\frac{1}{8} d^{cde} \Big[ (}  + f^{abc} (2a
\theta^{\rho\sigma} F^{d}_{\rho\sigma}F^{e\alpha\beta} +a
\theta^{\alpha\beta} F^{d}_{\rho\sigma}F^{e\rho\sigma}
\nonumber \\
& & \phantom{\frac{1}{8} d^{cde} \Big[ (}\left.+ 4
\theta_{\rho\sigma} F^{d\alpha\rho}F^{e\beta\sigma} + 8
\theta^{\alpha\rho} F^{d\beta\mu}{F^e_{\mu\rho}})\right].
\label{T4}
\end{eqnarray}
The divergent parts are calculated in the momentum representation
via dimensional regularization, by picking relevant terms out of
the expansion (\ref{trlog}). The resulting contributions are given
by
\begin{eqnarray}
{\mathrm{D}_1}^{\mathrm{div}} & =&\frac{\mathrm{i}}{2}
{\mathrm{Tr}} \left( \left(\Box^{-1} N_1\right)^2 \left(\Box^{-1}
T_4\right)\right)^{\mathrm{div}}
\label{D1} \\
&=& \frac{h}{(4\pi)^2\epsilon}  d^{abc} \int {\mathrm{d}}^4 x
\left[\frac{a-3}{4}(\theta^{\alpha\mu} F^a_{\alpha\nu} +
\theta_{\alpha\nu} F^{a\alpha\mu} ) (V_\mu V_\rho V^\rho V^\nu
)^{bc} \phantom{\frac{1}{2}} \right.
\nonumber \\
& & \phantom{\frac{{\mathrm{i}}}{(4\pi)^2} \frac{1}{\epsilon}
d^{abc} \int {\mathrm{d}}^4  } \left. +\frac{3a-4}{4}
\theta^{\alpha\beta}F^a_{\alpha\beta}(V^\mu V^\nu V_\nu
V_\mu)^{bc}\right],
\nonumber \\
{\mathrm{D}_2}^{\mathrm{div}} & =&  -\frac{\mathrm{i}}{2}
{\mathrm{Tr}} \left( \left(\Box^{-1} N_1\right)^3 \left(\Box^{-1}
T_3\right)\right)^{\mathrm{div}}
\label{D2} \\
& = & \frac{h}{(4\pi)^2\epsilon}  d^{abc} \int {\mathrm{d}}^4 x
\left[ \frac{7-3a}{6} \theta^{\alpha\beta} F^a_{\alpha\beta}
(V_\mu V^\mu V_\nu V^\nu + V_\mu V_\nu V^\mu V^\nu  \right.
\nonumber \\
& & \phantom{\frac{1}{(4\pi)^2\epsilon}  d^{abc} \int
{\mathrm{d}}^4 } + V_\mu V_\nu V^\nu V^\mu)^{bc} + \frac{3-2a}{6}
(\theta^{\alpha\mu} F^a_{\alpha\nu} \nonumber \\
& & \phantom{\frac{1}{(4\pi)^2\epsilon}  d^{abc} \int
{\mathrm{d}}^4 } + \theta_{\alpha\nu} F^{a\alpha\mu}) (V_\mu V^\nu
V^\rho V_\rho + V_\mu V_\rho V^\nu V^\rho
\nonumber \\
& & \phantom{\frac{1}{(4\pi)^2\epsilon}  d^{abc} \int
{\mathrm{d}}^4 }\left.\phantom{\frac{7a}{12} \theta
F^{a\alpha\mu}} + V_\mu V^\rho V_\rho V^\nu)^{bc} \right],
\nonumber \\
{\mathrm{D}_3}^{\mathrm{div}} &=& \frac{\mathrm{i}}{2}
{\mathrm{Tr}} \left( \left(\Box^{-1} N_1\right)^4 \left(\Box^{-1}
T_2\right)\right)^{\mathrm{div}}
\label{D3} \\
&=& \frac{h}{(4\pi)^2 \epsilon} d^{abc} \int {\mathrm{d}}^4 x
\left[ \frac{7a-11}{12} \theta^{\alpha\beta} F^a_{\alpha\beta}
(V_\mu V^\mu V_\nu V^\nu + V_\mu V_\nu V^\mu V^\nu  \right.
\nonumber \\
&  & \phantom{\frac{1}{(4\pi)^2 \epsilon} d^{abc} \int
{\mathrm{d}}^4}  + V_\mu V_\nu V^\nu V^\mu)^{bc} +\frac{a-3}{12}
(\theta^{\alpha\mu} F^a_{\alpha\nu} \nonumber \\
& & \phantom{\frac{1}{(4\pi)^2\epsilon}  d^{abc} \int
{\mathrm{d}}^4 } + \theta_{\alpha\nu} F^{a\alpha\mu}) (2 V^\rho
V_\rho V_\mu V^\nu + 2 V^\rho V_\mu V_\rho V^\mu
\nonumber \\
&&  \phantom{\frac{1}{(4\pi)^2\epsilon}  d^{abc} \int
{\mathrm{d}}^4 }\left.\phantom{\frac{7a}{12} F^{a\alpha\mu} )\ }+
V^\rho V_\mu V^\nu V_\rho +V_\mu V^\rho V_\rho V^\nu )^{bc}
\right],
\nonumber \\
{\mathrm{D}_4}^{\mathrm{div}} &=& -\frac{\mathrm{i}}{2}
{\mathrm{Tr}} \left( \left(\Box^{-1} N_2\right) \left(\Box^{-1}
T_4\right)\right)^{\mathrm{div}}
\label{D4} \\
&=& \frac{h}{(4\pi)^2 \epsilon} d^{abc} \int {\mathrm{d}}^4 x
\left[ \frac{4-3a}{4} \theta^{\alpha\beta} F^a_{\alpha\beta}
(V_\mu V_\nu V^\nu V^\mu)^{bc} \right. \nonumber \\ & &
\phantom{\frac{h}{(4\pi)^2 \epsilon} d^{abc} \int {\mathrm{d}}^4 }
+ \frac{2-a}{2} (\theta^{\alpha\mu} F^a_{\alpha\nu} +
\theta_{\alpha\nu} F^{a\alpha\mu}) (V_\mu V^\rho V_\rho
V^\nu)^{bc} \nonumber \\ & & \phantom{\frac{h}{(4\pi)^2 \epsilon}
d^{abc} \int {\mathrm{d}}^4 } + 2 (a+1) {\mathrm{i}}
\theta_{\alpha\nu}
F^a_{\beta\mu} (V^\mu F^{\alpha\beta} V^\nu)^{bc} \nonumber \\
& & \phantom{\frac{h}{(4\pi)^2 \epsilon} d^{abc} \int
{\mathrm{d}}^4 }+ \frac{2-a}{2} \mathrm{i} \theta^{\mu\nu}
F^a_{\mu\nu} (V^\alpha F_{\alpha\beta} V^\beta)^{bc}
\nonumber \\
& & \phantom{\frac{h}{(4\pi)^2 \epsilon} d^{abc} \int
{\mathrm{d}}^4 }  + 2{\mathrm{i}} \theta^{\alpha\beta}
F^a_{\beta\mu} (V_\nu F^{\mu\nu} V_\alpha- V^\mu F_{\alpha\nu}
V^\nu)^{bc}
\nonumber \\
& & \phantom{\frac{h}{(4\pi)^2 \epsilon} d^{abc} \int
{\mathrm{d}}^4 }  + {\mathrm{i}} \theta^{\alpha\beta} F^a_{\mu\nu}
(V^\mu F_{\alpha\beta} V^\nu- V_\alpha F^{\mu\nu} V_\beta)^{bc}
\nonumber \\
& & \phantom{\frac{h}{(4\pi)^2 \epsilon} d^{abc} \int
{\mathrm{d}}^4 }  - 2 N \theta^{\beta\mu} F^a_{\mu\nu}
F^{b\alpha\nu} F^{c}_{\alpha\beta}+ N \theta^{\mu\nu}
F^a_{\alpha\mu} F^b_{\beta\nu} F^{c\alpha\beta}
\nonumber \\
& & \phantom{\frac{h}{(4\pi)^2 \epsilon} d^{abc} \int
{\mathrm{d}}^4 } \left. -a\frac{3N}{4} \theta^{\mu\nu}
F^a_{\mu\nu} F^b_{\alpha\beta} F^{c\alpha\beta}   \right] ,
\nonumber \\
{\mathrm{D}_5}^{\mathrm{div}} &=& \frac{\mathrm{i}}{2}
{\mathrm{Tr}} \left( \left(\Box^{-1} N_2\right)^2 \left(\Box^{-1}
T_2\right)\right)^{\mathrm{div}}\\ &=& \frac{h}{(4\pi)^2 \epsilon}
d^{abc} \int {\mathrm{d}}^4 x \left[ \frac{a-3}{2}
(\theta^{\alpha\mu} F^{a}_{\alpha\nu} + \theta_{\alpha\nu}
F^{a\alpha\mu}) (F_{\mu\rho}F^{\nu\rho})^{bc}
\phantom{\frac{1}{4}} \right. \label{D5}
\nonumber\\
& & \phantom{\frac{h}{(4\pi)^2 \epsilon} d^{abc} \int
{\mathrm{d}}^4 }  + \frac{3a-4}{4}
\theta^{\alpha\beta}F^a_{\alpha\beta} (F_{\mu\nu}
F^{\mu\nu})^{bc} \nonumber\\
& & \phantom{\frac{h}{(4\pi)^2 \epsilon} d^{abc} \int
{\mathrm{d}}^4 } \left.+ \frac{4a-7}{4}
\theta^{\alpha\beta}F^a_{\alpha\beta} (V^\mu V_\mu V^\nu
V_\nu)^{bc} \right],
\nonumber \\
{\mathrm{D}_6}^{\mathrm{div}} &=& \frac{\mathrm{i}}{2}
{\mathrm{Tr}} \left( \left(\Box^{-1} N_1\right) \left(\Box^{-1}
N_2\right) \left(\Box^{-1} T_3\right) \right. \label{D6} \\ & &
\phantom{\frac{\mathrm{i}}{2} {\mathrm{Tr}}} \left. +
\left(\Box^{-1} N_2\right) \left(\Box^{-1} N_1\right)
\left(\Box^{-1} T_3\right)\right)^{\mathrm{div}} \nonumber
\\
&=& \frac{h}{(4\pi)^2 \epsilon} d^{abc} \int \mathrm{d}^4 x \left[
\frac{a-3}{2}\theta^{\alpha\mu} F^a_{\alpha\nu} (2 V_\mu V_\rho
V^\rho V^\nu + V_\rho V^\rho V_\mu V^\nu   \right.
\nonumber\\
& & \phantom{\frac{1}{(4\pi)^2 \epsilon} d^{abc} \int \mathrm{d}^4
}\phantom{\frac{a-3}{2}\theta^{\alpha\mu} F^a_{\alpha} }+V_\rho
V^\rho V^\nu V_\mu)^{bc}
\nonumber\\
& & \phantom{\frac{1}{(4\pi)^2 \epsilon} d^{abc} \int \mathrm{d}
^4 } +\frac{5a-4}{4} \theta^{\alpha\beta}F^a_{\alpha\beta} (V_\mu
V^\mu V^\nu V_\nu )^{bc}
\nonumber\\
& & \phantom{\frac{1}{(4\pi)^2 \epsilon} d^{abc} \int \mathrm{d}
^4 } +\frac{3a-4}{4} \theta^{\alpha\beta}F^a_{\alpha\beta} (V_\mu
V^\nu V_\nu  V^\mu)^{bc}
\nonumber\\
& & \phantom{\frac{1}{(4\pi)^2 \epsilon} d^{abc} \int \mathrm{d}
^4 }  -\frac{a}{2} \theta^{\alpha\beta}F^a_{\alpha\beta} (V_\mu
V^\nu V^\mu V_\nu)^{bc} -\frac{a}{4}
\theta^{\alpha\beta}F^a_{\alpha\beta} (F^{\mu\nu} F_{\mu\nu})^{bc}
\nonumber\\
& & \phantom{\frac{1}{(4\pi)^2 \epsilon} d^{abc} \int \mathrm{d}
^4 }  -2 \mathrm{i} (a+1) \theta_{\alpha\beta}F^a_{\mu\nu} (V^\mu
F^{\beta\nu} V^\alpha )^{bc}
\nonumber\\
& & \phantom{\frac{1}{(4\pi)^2 \epsilon} d^{abc} \int \mathrm{d}
^4 } + (a+1) \theta_{\alpha\beta}F^a_{\mu\nu} (F^{\alpha\nu}
F^{\beta\mu})^{bc} \nonumber\\
& & \phantom{\frac{1}{(4\pi)^2 \epsilon} d^{abc} \int \mathrm{d}
^4 } -2\mathrm{i} \theta^{\alpha\mu} F^a_{\alpha\nu} (V^\nu
F_{\mu\rho} V^\rho)^{bc} +2 \mathrm{i} \theta^{\alpha\mu}
F^a_{\alpha\nu} (V_\rho F^{\nu\rho} V_\mu)^{bc} \nonumber\\
& & \phantom{\frac{1}{(4\pi)^2 \epsilon} d^{abc} \int \mathrm{d}
^4 } + \theta^{\alpha\mu} F^a_{\alpha\nu} (F^{\nu\rho}
F_{\mu\rho})^{bc} + \theta^{\alpha\mu} F^a_{\alpha\nu}
(F_{\mu\rho} F^{\nu\rho})^{bc} \nonumber\\
& & \phantom{\frac{1}{(4\pi)^2 \epsilon} d^{abc} \int \mathrm{d}
^4} \left.+\frac{\mathrm{i}}{2} \theta^{\alpha\beta}
F^a_{\alpha\beta} (V^\mu F_{\mu\nu} V^\nu)^{bc} - \mathrm{i}
\theta^{\alpha\beta} F^a_{\mu\nu} (V^\mu F_{\alpha\beta}
V^\nu)^{bc} \right],
\nonumber \\
{{\mathrm{D}_7}^{\mathrm{div}}} &=&
\label{D7}-\frac{\mathrm{i}}{2} {\mathrm{Tr}} \left( \left(
\left(\Box^{-1} N_1\right)^2 \left(\Box^{-1} N_2\right)
\left(\Box^{-1} T_2\right)\right)^{\mathrm{div}} \right.
\\ && \phantom{-\frac{\mathrm{i}}{2} {\mathrm{Tr}}}
+\left( \left(\Box^{-1} N_1\right)^2 \left(\Box^{-1} T_2\right)
\left(\Box^{-1} N_2\right)\right)^{\mathrm{div}}  \nonumber
\\ &&
\phantom{-\frac{\mathrm{i}}{2} {\mathrm{Tr}}}+ \left.\left(
\left(\Box^{-1} N_1\right) \left(\Box^{-1} T_2\right)
\left(\Box^{-1} N_1\right)
\left(\Box^{-1} N_2\right)\right)^{\mathrm{div}} \right) \nonumber\\
&= &\frac{h}{(4\pi)^2 \epsilon} d^{abc} \int \mathrm{d}^4 x \left[
\frac{18-11a}{12} \theta^{\alpha\beta} F^a_{\alpha\beta} (2V^\mu
V_\mu V^\nu V_\nu + V^\mu V^\nu V_\nu V_\mu)^{bc}  \right.
\nonumber\\
& & \phantom{\frac{h}{(4\pi)^2 \epsilon} d^{abc} \int \mathrm{d}^4
} +\frac{3-a}{6} (\theta^{\alpha\mu} F^a_{\alpha\nu} +
\theta_{\alpha\nu} F^{a\alpha\mu}) (V_\mu V^\nu V^\rho V_\rho
\nonumber\\
& & \phantom{\frac{h}{(4\pi)^2 \epsilon} d^{abc} \int \mathrm{d}^4
} \left.\phantom{+\frac{3-a}{6} (}+ V_\mu V^\rho V_\rho V^\nu +
V_\mu V^\rho V_\rho V^\nu)^{bc} \right]. \nonumber
\end{eqnarray}
Their sum, that is the one-loop divergent part of the gauge boson
interaction
\begin{eqnarray}
\sum_{i=1}^7 {\mathrm{D}_i}^{\mathrm{div}} &=& \frac{N}{(4\pi)^2
\epsilon} h \theta^{\mu\nu} d^{abc} \int \mathrm{d}^4x \left(
-\frac{25a-3 }{48} F^a_{\mu\nu} F^b_{\rho\sigma}F^{c\rho\sigma}
\right. \nonumber \\
& &\phantom{\frac{N}{(4\pi)^2 \epsilon} h \theta^{\mu\nu} d^{abc}
\int \mathrm{d}^4x \Big(} \left.+\frac{a+21}{12}\, F^a_{\mu\rho}
F^b_{\nu\sigma} F^{c\rho\sigma} \right), \label{div}
\end{eqnarray}
has been obtained after the dimensional regularization and
summation of all the contributions.  Therefore, the total
divergent contribution to the effective action (\ref{m}) is
\begin{eqnarray}
\Gamma^{\mathrm{div}} &=&
\frac{11}{6}\frac{N}{(4\pi)^2\epsilon}\int \mathrm{d}^4x
F_{\m\n}^a F^{\m\n a}\label{diva} \\& &+\frac{N}{(4\pi)^2
\epsilon} h \theta^{\mu\nu} d^{abc} \int \mathrm{d}^4x \left(
-\frac{25a-3 }{48} F^a_{\mu\nu} F^b_{\rho\sigma} + \frac{a+21}{12}
F^a_{\mu\rho} F^b_{\nu\sigma} \right)F^{c\rho\sigma} . \nonumber
\end{eqnarray}
In the above expression the ghost contribution to the one-loop
effective action is included.

We are interested in the
renormalization of the theory. Our starting Lagrangian in
$D=4-\epsilon$ dimensional space-time has the following form:
\begin{eqnarray}
\mathcal{L}&=&-\frac{1}{4} F^a_{\mu\nu} F^{a\mu\nu} + \frac{1}{4}
g\mu^{\epsilon/2}h \theta^{\mu\nu} d^{abc}\left(
\frac{a}{4}F^a_{\mu\nu} F^b_{\rho\sigma} -F^a_{\mu\rho}
F^b_{\nu\sigma}\right)F^{c\rho\sigma} , \label{L}
\end{eqnarray}
where $g$ is the gauge coupling constant and $\mu$ is the
subtraction point mass parameter or the so-called renormalization
point. In order to cancel divergences, counter terms should be
added to the starting action, which produces the bare Lagrangian
from (\ref{diva}):
\begin{eqnarray}
\mathcal{L}_0&=&-\frac{1}{4} F^a_{\mu\nu} F^{a\mu\nu}
-\frac{11Ng^2}{6(4\pi)^2\epsilon}F^a_{\mu\nu} F^{a\mu\nu}
\nonumber\\& & +\frac{1}{4}g\mu^{\epsilon/2} h \theta^{\mu\nu}
d^{abc}\left( \frac{a}{4}F^a_{\mu\nu} F^b_{\rho\sigma}
-F^a_{\mu\rho}F^b_{\nu\sigma} \right)F^{c\rho\sigma}
\nonumber\\
& & -\frac{Ng^3\mu^{\epsilon/2}}{(4\pi)^2\epsilon} h
\theta^{\mu\nu} d^{abc}\left(-\frac{25a-3}{48}F^a_{\mu\nu}
F^b_{\rho\sigma} +\frac{a+21}{12}F^a_{\mu\rho} F^b_{\nu\sigma}
\right)F^{c\rho\sigma}
\nonumber\\
&=&-\frac{1}{4} {F_0}^a_{\mu\nu}{F_0}^{a\mu\nu} +\frac{1}{4}
g\mu^{\epsilon/2} h \theta^{\mu\nu}d^{abc} \left[\frac{a}{4}
\left(1+\frac{25a-3}{3a}\frac{Ng^2}{(4\pi)^2\epsilon}\right)
F^a_{\mu\nu} F^b_{\rho\sigma}\right.\nonumber
\\
& & \phantom{-\frac{1}{4} {F_0}^a_{\mu\nu}{F_0}^{a\mu\nu} } -
\left.\left(1+\frac{21+a}{3}\frac{Ng^2}{(4\pi)^2\epsilon}\right)
F^a_{\mu\rho}
F^b_{\nu\sigma} \right]F^{c\rho\sigma}. \label{LNbare}
\end{eqnarray}
It is easy to see that in order to keep the ratio of  the
coefficients of  two terms from (\ref{LNbare}) the same as in the
classical action (\ref{m}), one has to impose the condition
\begin{equation}
\left(-\frac{25a-3 }{48}\right):\left(\frac{a+21}{12}\right) =
\frac{a}{4} : (-1). \label{cond}
\end{equation}
Interestingly enough, this equation has two solutions, $a=1$ and
$a=3$. In our previous paper \cite{Buric:2005xe} the action
(\ref{m}) was discussed and renormalizability was proved for
$a=1$. In this case, the divergences are canceled through
redefinition of the gauge potential and the coupling constant.

The case $a=3$ is different since the non-commutative deformation
parameter $h$ has to
be renormalized. The bare gauge field, the coupling and the NC
deformation parameter are defined as follows:
\begin{eqnarray}
V^\mu_0&=&V^\mu\sqrt{1+\frac{22Ng^2}{3(4\pi)^2 \epsilon}},
\label{V}\\
g_0&=&\frac{g\mu^{\epsilon/2}}{\sqrt{1+\frac{22Ng^2}{3(4\pi)^2
\epsilon}}},
\label{g}\\
h_0&=&\frac{h}{1-\frac{2Ng^2}{3(4\pi)^2\epsilon}}, \label{h}
\end{eqnarray}
with an arbitrary choice for the renormalization point $\mu$. Any
change in $\mu$ is compensated by the corresponding change in the
charge $g$, the NC deformation parameter $h$ and for the scale of the
fields. The above result  means that it is not possible to
renormalizes our action, for $a=3$, only through the
renormalization of the vector potential and the coupling constant.

\initiate\section{Ultraviolet asymptotic behavior of NC SU(N) gauge
theory}

In this section we investigate the high-energy behavior of our theory
(\ref{m}) by employing the renormalization group equation (RGE)
and compute the relevant $\beta$ functions. The RGE provides a
framework within which we discuss the ultraviolet (UV) asymptotic
behavior of renormalizable gauge field theory (GFT), i.e. the
behavior of the relevant amplitudes in a physically uninteresting
region, i.e. in a region for large $g$ and/or far from the origin.

Since the gauge coupling constant $g$ in our theory (\ref{m})
depends on the renormalization point $\mu$ {\it satisfying the
same beta function}
\begin{equation}
\beta_g=\mu \frac{\partial }{\partial\mu}g(\mu)
=-\frac{11Ng^3(\mu)}{3(4\pi)^2}, \label{betag}
\end{equation}
as for the commutative Yang-Mills theory without fermions and with
gauge independence in lowest order ($g^3$), {\it our theory is} UV
{\it stable}. This means that (\ref{m}) belongs to the class of
asymptotically approaching free-field theories, or in short `{\it
asymptotically free theory}'. The solution to (\ref{betag}) is the
very well-known result \cite{Gross:1973id,Weinberg:1996kr}
\begin{equation}
\alpha_s(\mu)=\frac{g^2(\mu)}{4\pi}=
\frac{6\pi}{11N}\frac{1}{\ln\frac{\mu}{\Lambda}}. \label{alphas}
\end{equation}
In (\ref{alphas}), $\Lambda$ is an integration constant not
predicted by the theory, thus it is a free parameter to be
determined from the experiment. The QCD (physical) interpretation
of $\Lambda$ is that it represents the marking of the boundary
between a world of quasi-free quarks and gluons and the world of
protons, pions, and so on. For typical QCD energies $\mu$ with
$m_b \ll \mu \ll m_t$, where fermions are included ($N_f=5$), the
study of hadronic production in $e^+e^-$ annihilation at the Z
resonance has given a direct measured value $\alpha_s(m_Z)=0.12$
corresponding to $\Lambda= \Lambda_{\rm QCD}\simeq 250$ MeV.

The $\beta$ function for the NC deformation parameter $h$ can be
easily computed from (\ref{h}) and (\ref{betag}):
\begin{equation}
\beta_h=\mu \frac{\partial}{\partial\mu}\,h(\mu)
=-\frac{11Ng^2(\mu)}{24\pi^2}\,h(\mu).
\label{betah}
\end{equation}
Since both $\beta$ functions (\ref{betag}) and (\ref{betah}) are
{\it negative}, both the coupling $g$ and the NC deformation
parameter $h$ decrease with increasing energy and our theory is
considered to be UV stable.

Solving equation (\ref{betah}) we obtain
\begin{equation}
h(\m) = \frac{h_0}{\ln\frac{\mu}{\Lambda}},
\label{H1}
\end{equation}
which is  the {\it running NC deformation  parameter $h$}. Here
$h_0$ is an additional integration constant whose physical
interpretation is going to be discussed later. From the above
expression we see that with the increase of energy the NC
deformation parameter decreases, which might seem counterintuitive
in the view of Heisenberg uncertainty relations. However, there
are many arguments for the modification of uncertainty relations
at high energy \cite{gma}. For example, if the commutation
relation is\footnote{The physical effects of the modifications at
hight energies, like (\ref{He}), in 1 and 3 dimensions were
analyzed in \cite{Chang:2001kn,Benczik:2005bh} and references
therein.}
\begin{equation}
[x,p]=i\hbar(1+\beta p^2),
\label{He}
\end{equation}
where $\beta$ is a constant and has dimension $energy^{-2}$, then
one can easily see  that in the region of the large momentum
$\Delta x$ grows linearly \cite{kmm},
\begin{equation}
\Delta x=\frac{\hbar}{2}\left(\frac{1}{\Delta p}+\beta \Delta
p\right).
\label{HD}
\end{equation}
From this example it follows that large energies do not
necessarily correspond to small distances, that is the behavior of
the running NC deformation parameter (\ref{h}) does not imply that
non-commutativity vanishes at small distances. This is related to
the UV/IR correspondence.

Owing to the necessity of the renormalization of the
non-commutativity deformation parameter $h$, the scale of
non-commutativity $\Lambda_{\rm NC}$ becomes a function of energy
$\mu$ too,
\begin{equation}
\Lambda_{\rm NC}(\mu) =\Lambda_{\theta}\,\sqrt{\ln
\frac{\mu}{\Lambda}}\ . \label{lambdar}
\end{equation}

Equivalently to (\ref{H1}), the scale $\Lambda_{\rm NC}$ becomes the
{\it running scale of non-commu\-ta\-ti\-vi\-ty}. Here
$\Lambda_{\theta}$ is an additional integration constant, namely the
dimension of energy, not predicted by the theory.

Even though that the physical interpretation of $h_0$ and/or
$\Lambda_{\theta}$ is not quite clear, it seems that, owing to the
energy dependence, they have to be proportional to the scale of
non-commutativity $\Lambda_{\rm NC}$. If one could think of $h_0$
and/or $\Lambda_{\theta}$ as a boundary between a world of
commutative fields (particles) and non-commutative quantum fields,
then, according to (\ref{H1}) and (\ref{lambdar}), it would be
obvious to assume that in a first approximation
$h_0=1/\Lambda^2_{\theta}=1/\Lambda^2_{\rm NC}$. Considering
typical QCD energies, the factor $\sqrt{\ln({m_Z}/{\Lambda_{\rm
QCD}})}\simeq 2.4 $, which means that at such energies the scale
of non-commutativity $\Lambda_{\rm NC}$ is effectively shifted by
a factor of $\simeq 2.4$ up.

\initiate\section{Discussion and conclusion}

 We have constructed a version of the SU(N) model on
non-commutative Minkowski space at first order in
the non-commutative deformation parameter $h$,
which has the one-loop multiplicative
renormalizable gauge sector.

We have shown in \cite{Buric:2005xe} that if the gauge fields are
in the adjoint representation of SU(N), the action (\ref{actioa})
for the freedom parameter $a=1$ is renormalizable. Trying to
extend this result to the gauge group of the standard model $\rm
U(1)_Y \otimes SU(2)_L\otimes SU(3)_C$, we have seen
\cite{Buric:2006wm} that the action of the type  (\ref{action})
with SW map (\ref{fields1}) and/or, (\ref{actioa}), for $a=1$
cannot be renormalized. However, with a suitable choice of the
representations of the gauge group, the theory is renormalizable
and finite for the freedom parameter $a=3$ \cite{Buric:2006wm}.
This naturally poses a question: is the obtained result, $a=3$,
just an outcome of a specific interaction among gauge bosons in
the NCSM or is there something new?

In order to answer the above question,
in this paper  we reconsider
the renormalizability of the `building blocks'
for arbitrary values of the freedom parameter $a$
i.e. of the
non-commutative SU(N) gauge theories described by the action (\ref{m})
and for the gauge fields in the adjoint representation.

As a solution to the problem of the origin of the freedom
parameter $a$, we propose the framework where the general
non-commutative action was expanded in terms of NC field
strengths, with the $\star$-product defining multiplication. This
way the higher order non-commutative gauge interaction was
introduced and higher derivative interaction terms have been
constructed: $(\,b\,h\,\theta^{\mu\nu}\widehat
F_{\mu\nu}\star\widehat F_{\rho\sigma}\star\widehat
F^{\rho\sigma})$ and $(\,c\,h\,\theta^{\mu\nu}\widehat
F_{\mu\rho}\star\widehat F_{\nu\sigma} \star\widehat
F^{\rho\sigma})$. Seiberg-Witten mapping (\ref{fields1}) and
freedom in constants $b$ and $c$, ($b \not= 0, c=0$), lead to
relation between $a$ and $b$: ($a=1+4b$). Thus, in the above
defined framework, the NC SU(N) gauge field  theory is described
by the NC action,
\begin{eqnarray}
S^a_\star &=&{\rm Tr}\int \mathrm{d}^4x \left(-\frac{1}{2}
\widehat F_{\mu\nu} \star \widehat F^{\mu\nu} + \frac{a-1}{4}\, h
\, \theta^{\mu\nu} \widehat F_{\mu\nu}\star\widehat F_{\rho\sigma}
\star\widehat F^{\rho\sigma}\right), \label{GenAct}
\end{eqnarray}
which, together with well known Seiberg-Witten mapping, (\ref{fields1}),
leads to the action in terms of commutative fields, (\ref{actioa}),
that is one-loop renormalizable only for $a=1,3$ and for various
representations of the gauge potential,
(\cite{Buric:2005xe,Buric:2006wm,Martin:2006gw}, and this work). Those
results suggest that in the further investigations of the
renormalizability properties, of the $\theta$-expanded
non-commutative gauge field theories involving fermion and Higgs
fields, the gauge action (\ref{GenAct}) should be used as a
starting point.
Factor $a$ from (\ref{GenAct}) and/or (\ref{actioa}) can be kept as the
freedom parameter generically, i.e. during the computations.

To obtain 1-loop renormalizable NC SU(N) gauge theory for $a=3$
we had to pay a prize by the renormalization of
the non-commutative deformation parameter $h$.
However, by doing that we gain the beta function $\beta_h$
which allow us to analyze the ultraviolet behavior
of the parameter $h$.

The analysis of the
ultraviolet asymptotic behavior of NC SU(N) gauge theory
in the case $a=3$ is in order next. The
relevant beta functions, $\beta_g$ and $\beta_h$, have been
computed and they are both negative, thus
{\it our theory is completely} UV {\it stable}. The non-commutative
deformation parameter $h$ becomes {\it the running
non-commutative deformation
parameter and it is asymptotically free}. However, owing to the
inverse square behavior, {\it the non-commutative scale runs
according to} (\ref{lambdar}). The function (\ref{lambdar}) is
very smooth and mild, showing a small change of the scale of
non-commutativity as the energy increases. We consider this
property very welcome, because it shows a large degree of
stability of our theory within a wide range of energy.

The necessity of the $h$ renormalization jeopardizes the previous
hope that the NC SU(N) gauge theory might be renormalizable to all
orders in $h\theta^{\mu\nu}$. This means that most probably the
theory would need to be renormalized independently order by order
in the non-commutative parameter $h\theta^{\mu\nu}$.

The one-loop renormalizability of the non-commutative SU(N) gauge
sector is certainly a very encouraging result, both theoretically
and experimentally. So far, this property has not concerned
fermions: the results on the renormalizability of NC theories
including the Dirac fermions are not yet positive,
\cite{Wulkenhaar:2001sq,Maja}, i.e. till now fermions in
non-commutative theory have still spoilt the UV stable behavior of
(\ref{betag}) and (\ref{betah}) owing to non-renormalizability.

Non-Abelian commutative theories are completely renormalizable.
However, fermions spoil the stable behavior of a gauge boson beta
function, but they leave room to spare, and the theory becomes
asymptotically free as long as $\frac{11}{4}C_2(G)>T(R)$
\cite{Gross:1973id}. The possibility that something similar could
happen in the case of the NCSM
\cite{Buric:2006wm,Calmet:2001na,Blazenka,Aschieri:2002mc,Goran},
at first order in the NC parameter $h\theta^{\mu\nu}$, is a matter
for other, fermion involving, studies.

The present result has deep impact on $\theta$-expanded non-commutative gauge
field theory and it
could be an indication that the
inclusion of fermions into a renormalizable
theory might be possible within the framework of
some higher non-commutative gauge interaction symmetry,
similar to the scheme outlined by (\ref{GenAct}).
More careful choices of freedom
parameters, of the representations of a gauge group and of the
renormalization of the non-commutative parameter as well, are certainly necessary.

We hope that the answer to the obvious question,
`why the freedom parameter $a$ is so special?' has been well explained in this article.
Clearly, it led us to discovery of the key role of
the higher order non-commutative
gauge interaction in one-loop renormalization of the
$\theta$-expanded gauge theories, at first order in $h\theta$, and
hence, to discovery that the non-commutative deformation parameter
$h$ vanishes at very high energies, i.e. that $h$
is asymptotically free, opposite to the previous expectations.
Those discoveries are certainly of the paramount importance
for further research of the non-commutative gauge field theory properties.


\begin{ack}
We would like to thank M. Buri\' c, S. Doplicher, R. Horvat, T. Ohl,
J. Wess and R. Wulkenhaar for a very fruitful discussions. The work
of D.~L. and V.~R. is supported by the project 141036 of the Serbian
Ministry of Science. The work of J.~T. is supported by the project
098-0982930-2900 of the Croatian Ministry of Science Education and
Sports.
\end{ack}


\end{document}